\documentclass[11pt]{report}
\usepackage{color}
\usepackage{graphicx}

\begin{document}

\baselineskip=15 pt

\centerline{\Large {\bf On the mass distribution of neutron stars}}

\bigskip
\bigskip
\centerline{R.Valentim, E. Rangel and J.E. Horvath}
\bigskip
\centerline{Instituto de Astronomia, Geof\'\i sica e Ci\^encias
Atmosf\'ericas}

\centerline{Universidade de S\~ao Paulo, 05570-010 Cidade
Universit\'aria SP, Brazil}


\vspace{0.8 cm}

\noindent{\bf Abstract}

\bigskip{The distribution of masses for neutron stars is
analyzed using the Bayesian statistical inference, evaluating the
likelihood of proposed gaussian peaks by using fifty-four measured
points obtained in a variety of systems. The results strongly
suggest the existence of a bimodal distribution of the masses,
with the first peak around $1.37 {M_{\odot}}$, a much wider second
peak at $1.73 {M_{\odot}}$. The results support earlier views
related to the different evolutionary histories of the members for
the first two peaks, which produces a natural separation (even if
no attempt to ``label'' the systems has been made here), and
argues against the single-mass scale viewpoint. The bimodal
distribution can also accommodate the recent findings of $\sim
M_{\odot}$ masses quite naturally. Finally, we explore the
existence of a subgroup around $1.25 {M_{\odot}}$, finding weak,
if any, evidence for it. This recently claimed low-mass subgroup,
possibly related to $O-Mg-Ne$ core collapse events, has a
monotonically decreasing likelihood and does not stand out clearly
from the rest of the sample.}
\bigskip

{\bf Keywords} {Stars: neutron}

\vspace{0.4cm}

\noindent{\bf 1. Introduction}

\vspace{0.4cm}

The measurements of masses and radii of neutron stars have the
potential to constrain stellar evolution and dense matter physics
alike. In fact, matching stellar evolution results and arguments
to actual neutron star observations is crucial to test the whole
theory. As is well-known, he standard Stellar Evolution theory
suggests the mass at the main sequence to be the most important
parameter for the final outcome of massive stars. The lowest end
of $\sim 8-11 M_{\odot}$ is expected to produce very degenerate
$O-Mg-Ne$ cores which eventually collapse because of electron captures.
Some possible systems in which the collapse can occur have been
discussed in Siess (2007), Poelarends et al. (2008) , Nomoto \& Kondo (1991)
Nomoto \& Iben (1985) and Nomoto (1987), among other works.
Podsiadlowski et al. (2005) elaborated on this problem and suggested that,
due to the ``characteristic mass'' expected for the core, the equation of
state and amount of ejected mass would produce low-mass neutron stars in
the ballpark of $\sim 1.25 M_{\odot}$, also expected to receive
small natal kicks in their birth events and therefore showing a low
eccentricity. Further evidence in favor of this
proposal has been presented by Schwab, Podsiadlowski \& Rappaport (2010)
after analyzing a sample of 14 well-measured neutron stars. In addition,
a group of $\sim 1.35 M_{\odot}$ has been also identified and associated with
the standard scenario of iron core collapse.This group features a much higher
natal kick, and comprises some of the binary pulsar systems such as PSR 1913+16.

On the other hand, increasing evidence for massive neutron stars has
been mounting, with several systems in the range $\sim 1.6-1.8 M_{\odot}$ and
the very recent report of a $ 1.97 \pm 0.04 M_{\odot}$ (Demorest et al. 2010).
Interestingly enough, most of these neutron stars typically have a
white dwarf companion, and thus a distinct evolutionary history. However, systems
in HMXRB and in a binary pulsar also exist. The mass could be the direct result of
the existence of massive iron cores for $M \geq 19 M_{\odot}$ in the Main Sequence
(Timmes, Woosley \& Weaver 1996), a view advocated by van den Heuvel (2010) and others.

The knowledge of this mass distribution is therefore fundamental
to understand the mechanisms involved in the final stages of
stellar evolution. In this work, we consider the sample of know
neutron star masses. We applied a Bayesian analysis for all set of
masses overcome an {\it a priori distribution}. This work is
motivated by a similar earlier approach by Finn (1994), who
applied the Bayesian approach for to estimate the upper and lower
limit for the neutron star masses, using one small set of data
with only four well-measured neutron star binary pulsar systems.
He observed the coincidence that all binary pulsar systems have
constrained the masses close to $1.35 {M\odot}$. Schwab,
Posialdowski \& Rappaport (2010) recently argued for the existence
of two distinct neutron stars populations, the first with
high-mass $\sim 1.35 {M_{\odot}}$ and the second with low-mass
$\sim 1.25 {M_{\odot}}$, in a work that analyzed fourteen
well-measured objects with uncertainties of $\leq 0.025
{M_{\odot}}$. They interpreted these two populations as to be
result of distinct evolutionary formation scenarios: low-mass
populations originates in electron-capture SNe and feature low
kicks, while the high-mass population is the result of iron core
collapse SNe.  Both the situations, the authors proposed that
masses are bounded the formation mechanisms restrict the range of
neutron stars and the existence of the two channels for the
production of neutron stars. Work by Thorsett and Chakrabarti
(1999) a decade ago concluded that a very narrow range of masses
around $\sim 1.35 {M_{\odot}}$ was found analyzing a sample of 50
objects, ruling out accretion at the level of $\Delta M \geq \,
0.1 M_{\odot}$ for the binary pulsars known at that time. Very
recently, Zhang et al. (2010) concluded that a substantial
accretion was present in recycled objects using an enlarged
sample. Clearly, a reanalysis of this subject is in order.

\vspace{0.4 cm}

\noindent{\bf 2.Neutron star sample}

\vspace{0.4 cm}

Our adopted sample is the compilation by Lattimer and
collaborators, publicly available at
http://stellarcollapse.org/nsmasses . After finding the central
values for each source (Fig. 1), we included the 55 neutron stars
with error bars varying between $0.009 {M_{\odot}}$ and $0.548
{M_{\odot}}$ in our work. The full references to the original
works, including the label letter employed in the compilation are
listed in the References. The uncertainties for each mass are
quite different because the methods of measurement were distinct
at different times, and the methods had been improved in many
cases.

\begin{figure}
\begin{center}
\includegraphics[width=10.0cm]{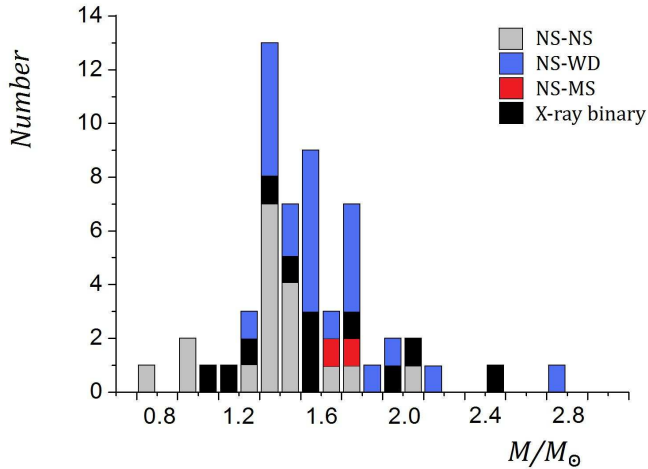}
\end{center}
\caption{Neutron Stars masses sample. The
corresponding uncertainties vary widely and have been included in
the analysis. Data taken from stellarcollapse.org.}
\end{figure}

\vspace{0.4cm}

\noindent{\bf 3. Statistical methodology}

\vspace{0.4cm}

In this work the analysis the sample of
neutron star masses has been studied using Bayesian
Statistics based on conditional probabilities, usually stated as
$p(H_i|D,I) = {{p(H_i|I)p(D|H_i,I)}\over{p(D|I)}}$, where $p(H_i|I)$
is the probability {\it  priori}, $p(D|H_i,I)$ is
the {\it likelihood function}, $p(H_i|D,I)$ is the {\it posteriori
probability} and $p(D|I)$ is the {\it predictive probability}.
This approach has been shown to be particularly
powerful for the treatment of
scarce/inaccurate data, yielding nonetheless quite accurate
estimate of the parameters in most cases.
One of the main tools to evaluate the quality of the results
is the {\it Bayesian Information Criterion} (BIC),
which considers the ratio of the likelihood models, the available
data and models, and explicitly penalizes the model having more parameters.
A crucial (and often criticized) ingredient is the {\it a priori}
expectations from theoretical inference which is weighed by the
BIC. An earlier study by
Finn (1994) used radio observations for four neutron star binary
pulsar systems and employed Bayesian
Statistics, approximating each observed point by a gaussian
function (on mass and standard deviation). He used one flat {\it a
priori} distribution between an assumed upper
limit of mass $m_u$ and the lower limit $m_l$. The values found were
$1.01<m_l/{M_{\odot}}<1.34$ (lower limit) and
$1.43<m_u/{M_{\odot}}<1.64$ (upper limit). We aimed to improve
this kind of analysis by working with the much larger sample
and exploiting the potential of the Bayesian formalism.

\vspace{0.4cm}

\noindent{\bf 3.1 Likelihood}

\vspace{0.4cm}

Our tasks in this section is to construct the likelihood function.
The likelihood distribution is the key point of Bayesian analysis
because it considers the data and the theoretical knowledge about
the measurements together. Here, we assumed that the likelihood function is
simply the product of independent probabilities for what was
measured and what was expected to be measured

$$
{\it L}(\theta|D,M) = \left[\prod_{i}^{N}
p(m_i|D,M)\right]\prod_{j} p(m_j|D,M).
$$

Where $\theta$ represents the space of parameters, ${D}$ is the
data set, ${M}$ the models, $p(m_i|D,M)$ is the probability of the
data to be measured and $p(m_j|D,M)$ is the expected probability for
the measured data. The likelihood weights the sampling probability
given the data ($D$) and assuming the model ($M$).
The likelihood in our case is

$$
{\it L}(\theta|D,M) = -\exp{\int
n_p(M,M_1,M_2,a_p,\sigma_1,\sigma_2,G) dM}
$$

$$
\prod{}{a}_p \times
{g}(M_1,\sigma_1,M_i,\xi_i)+(1-{a}_p-{a_{0}})\times
$$
$$
\times {g}(M_2,\sigma_2,M_i,\xi_i) + {a_{0}}\times
{g}(1.25,0.07,M_i,\xi_i)
$$

Where ${n}_p$ is a function that involves the peak masses $M_{1}$,
$M_{2}$, $a_p$ is the relative amplitude of the first peak,
${a_{0}}$\footnote{Where $0\leq a_p+{a_{0}}\leq 0$.} is the
amplitude centered on the $1.25 {M_{\odot}}$, with assumed
standard deviation $\sim 0.07$, $\sigma_1$, $\sigma_2$ are the
standard deviations of the theoretical peaks and ${g}$ is a
gaussian function (actually the product of two gaussian
distribution integrated).

$$
{g}= \int_{a_1}^{a_2}{\exp\left[{-{{(u-x_1)^2}\over{2
q_1^2}}}\right]} \times\exp\left[{-{{(u-x_2)^2}\over{2
q_2^2}}}\right]du
$$

Where $a_1$ and $a_2$ are

$$a_1 = max[x_1-H\times q_1, x_2-H\times q_2]$$

and

$$a_2 = max[x_1+H\times q_1, x_2+H\times q_2];$$

$H$ is the scale parameter, $q_1$ , $q_2$ are the standard
deviations of the $x_1$ and $x_2$. Finally, $n_p$ is

$$
n_p(M,M_1,M_2,a_p,\sigma_1,\sigma_2,g) = {a}_p \times
\exp\left[{-{{(M-M_1)^2}\over{2 s_1^2}}}\right]+
$$

$$
+{G}\times \exp\left[{-{{(M-1.25)^2}\over{2 (0.07^2)}}}\right]
+[1-({{a}_{p}}+a_{0})] \times\exp\left[{-{{(M-M_2)^2}\over{2
\sigma_1^2}}}\right]
$$

\vspace{0.4cm}

\noindent{\bf 3.2 The {\it a priori} distributions}

\vspace{0.4cm}

The {\it a priori} distribution is the assumed knowledge about the
phenomenon that is treated. In many previous works, notably those of
Finn (1994) and Schwab, Podsiadwolski and Rappaport (2010), gaussian
distributions were employed, and are a ``natural'' choice employed in our
work (see Finn 1994 for a discussion of this gaussian form within the Bayesian
approach). We then assumed, to make contact with the
work of Schwab, Podsiadwolski and Rappaport (2010), a distribution
peaked on two masses, around ${M}_1=1.35 {M_{\odot}}$, and
${M}_2=1.55 {M_{\odot}}$. Those authors restricted their analysis
to a set of 14 well-measured NS, and did not include the large error bar
points of the X-ray binaries and several WD-NS systems, consistently
with their frequentist approach. With the aim of exploring the full
distribution, we have chosen a second, higher value of ${M}_2$
to match the plain mean value of the NS in WD-NS and X-ray binary
systems of the sample first but, as we shall see below, the precise
value is not too important at this point. We
performed a first run within this two-value hypothesis, and compared it with
the occurrence of a {\it single} mass peak for all objects,
including NS-NS binaries, WD-NS binaries, X-ray binaries and MS-NS systems.
The motivation for this first bold comparison was to see whether a single
mass scale was still possible with the present data, as many works
insisted on till a few years ago.

After this ``first run'', and still working within the two-peak
hypothesis, and adopting the same values as Schwab, Podsiadwolski
and Rappaport (2010) for $M_{1}$ and $\sigma_{1}$, we looked for
evidence of a subgroup attributed by them to $O-Mg-Ne$ cores
producing low-mass neutron stars. In addition to the main peaks,
we examined the distribution to identify the presence of this
subgroup by defining ${a_{0}}$ as the peak amplitude of a $1.25
{M_{\odot}}$, that is, relative to the two main identified peaks.Again,
since the claimed masses lie at the low end, the calculation is not
sensitive to the precise value of $M_2$.
We defined and calculated the relevant likelihood of a third peak,
physically related to the formation of light NS out of $O-Mg-Ne$
cores of the lighter progenitors, of this second run.

In a third run we left the values of the masses $M_1$ and $M_2$ to
be determined directly by the raw available data, together with
their standard deviations values $\sigma_1$ and $\sigma_2$. That
is, while still imposing the distribution to be composed of two
gaussian forms, we sought for the optimal values without
restrictions as driven by the data sample. The purpose here is
to let the Bayesian tools to indicate which are the possible values
given the full error bars and within a definite gaussian hypothesis.

\vspace{0.4cm}

\noindent{\bf 4. Results and Conclusions}

\vspace{0.4cm}

We first address the basic results provided by the first run: the
likelihood of still a single mass-scale as an explanation of all
the data points is much lower than (at least) two peaks in them,
in spite of the introduction of extra parameters penalized by the
Bayesian approach. This results may not be meaningful for some,
since it has been known for years that the more massive systems
should have accreted $\sim 0.1-0.2 M_{\odot}$ or more, and
therefore detach from the original mass value. However, the first
run is important to overcome the idea that a single value of the
NS mass will be enough: even if the two peaks are quite close,
this distribution is preferred to a single wide peak. This is
somewhat expected, since the extremes of the determined masses,
around $1 M_{\odot}$, (van der Meer et al. 2007, see the
recent work by Rawls et al. 2011 released after the completion
of this work) at the lowest and $1.97\pm 0.04 M_{\odot}$
(Demorest et al. 2010), with a few more massive candidates
(Clark et al. 2002, Freire et al. 2008b), are
now separated by at least $1 M_{\odot}$.

\begin{figure}
\begin{center}
\includegraphics[width=8.0cm]{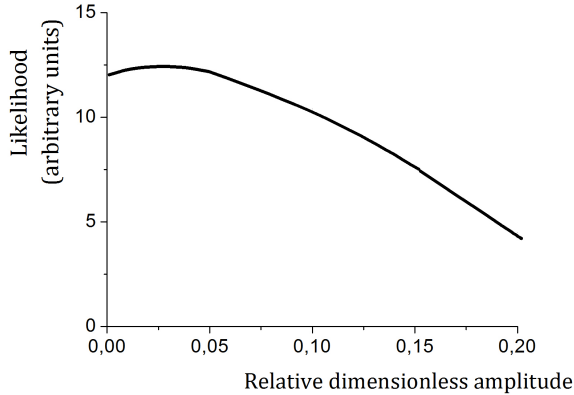}
\end{center}
\caption{The likelihood of a third,
low-mass peak. The existence of this third peak has a decreasing
likelihood except for a nearly constant value near $a_{0} \sim 0$}
\end{figure}

The results of the second run are summarized in Fig.2. The
calculated likelihood of the third peak is a monotonically falling
function of the amplitude, being maximal around zero amplitude
(relative to the prominent main peaks). Therefore, we conclude
that there is little evidence for the presence of a low-mass
subgroup, and that the four objects in this range may be in fact
members of the $1.35 M_{\odot}$ peak . However, the confirmation
of this result would have important consequences, since the
progenitor stars are quite abundant. The lack of strong evidence
of a peak $\sim 1.25 M_{\odot}$ could be due to still poor
sample/bias, or it could alternatively mean that most of the $8-11
M_{\odot}$ become AGB, an issue that deserves serious
consideration.

The results of the third run rendered two masses, as shown in
Fig.3. The first mass is not substantially different from the
result of Schwab, Podsiadwolski and Rappaport (2010), a fact we
interpret as the robustness of the distribution at this scale; while the
second one is now around $1.73 M_{\odot}$. This is not surprising,
since the known WD-NS and X-ray binary systems have large
uncertainties but also high values of the central NS masses. The
full advantage of the Bayesian techniques suggest here that masses
around $2 M_{\odot}$ are not unexpected, even less when
considering the obtained value of $\sigma_{2} = 0.25$ (for comparison, the value of
$\sigma_1 = 0.042$ is very narrow, as expected).
The recently announced value of the object in the
system, $ 1.97 \pm 0.04 M_{\odot}$ according to Demorest et al. 2010
(not included in the sample), is just an example of this
higher figure for the systems of this kind, and suggests a mean
accreted mass of several tenths of $M_{\odot}$, although the precise
value is model-dependent and should be estimated for each individual case.
This subgroup may also include
members coming from $\geq 20 M_{\odot}$ masses in the Main
Sequence (Timmes, Woosley and Weaver 1996, van den Heuvel 2004), born with
higher masses without suffering any substantial accretion..

\begin{figure}
\begin{center}
\includegraphics[width=6.0cm]{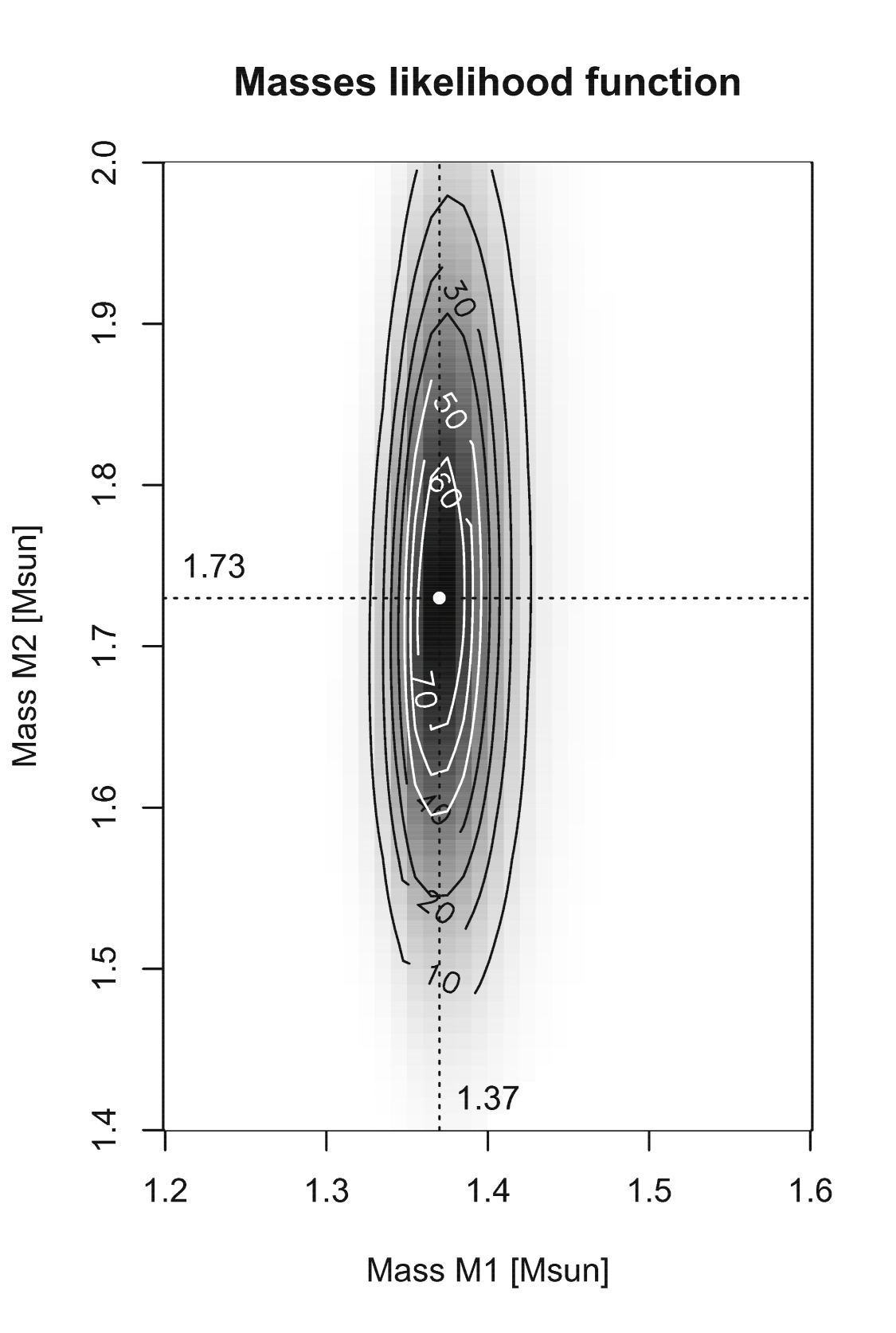}
\end{center}
\caption{The values of masses at
which the distribution peaks. The numbers on the contours
indicate the decreasing probabilities of finding $M_1$, $M_2$ there.}
\end{figure}

In summary, we have presented an analysis of an available NS mass
sample which indicates a) a bimodal distribution with a narrow
peak at $M_1 = 1.37 M_{\odot}$ (fully compatible with the findings
of Schwab, Podsiadwolski and Rappaport (2010), and a higher peak
at $M_1 = 1.73 M_{\odot}$, the latter with a wide shape capable of
accommodating $\geq 2 M_{\odot}$ masses, b) little evidence for
the expected low-mass NS descendants from the $8-11 M_{\odot}$
range in the Main Sequence, and c) the unsustainability of the
``one-mass-fits-all'' picture, since the adopted Bayesian scheme
properly weights the large uncertainties and extra parameters of
the bimodal hypothesis applied to an imperfect dataset. Given the
potential implications for the stellar evolution, the need of further
pursuing these kind of analysis and a continuous improvement by
incorporating new data can not be overstated.

After the completion of this work we noticed the public release of
a Bayesian analysis by Kiziltan, Kottas and Thorsett (2010), tackling the
same problem and with a very detailed treatment of the calibration. Also
their sample is restricted to avoid the inclusion of WD-NS and X-ray
binary systems with large error bars. Within this analysis, the authors
find two peaks, one at $M = 1.35 M_{\odot}$, and a second one located at
$M = 1.5 M_{\odot}$. Given the different sampling and their reliance on
simulated distributions (not performed by us), it is fair to say that
there is enough room for a convergence of results. It is important to note
that these authors warn against the inclusion of more uncertain points
because of a possible contamination of the sample. While we agree with
this judgement, we believe that unless the higher masses happen to be
completely and systematically overestimated, the emergence of a bump
at a mass even higher than $1.5 M_{\odot}$ is strongly expected. Our value
for that ``second peak'' should reflect to a large extent the difference
of including these loose observational data as processed by the Bayesian
formalism, acknowledged to handle this kind of situations better than
well-known frequentist approaches. In addition, the quest for a low-mass
bump remains in that work, since the peak at the lower $M = 1.35 M_{\odot}$
hosts the low-mass systems attributed by Schwab, Podsiadwolski and Rappaport
(2010) to the lower end of progenitor masses.

Finally, we acknowledge the works of Steiner, Lattimer and Brown (2010) and
Zhang et al. (2010), the first aimed to reveal the nuclear equation of state and
the second focusing on the evolutionary features of the systems.
Steiner, Lattimer and Brown found independently, from an analysis of
a subset of X-ray bursters,
that the maximum mass had to be quite high, as demanded by the 2-$M_{\odot}$
determination by Demorest et al. (2010) released shortly after their paper.
We did not attempt any specific
inference about the nuclear equation of state here, although our results also
demand a theoretical description capable to accommodate the objects in
the second peak. On the other hand, Zhang et al. used almost the same
sample, but their calculations attempted to link the period to the mass,
rather than discriminating between single-peak and multimodal distributions.
Even their statements about the NS-NS systems alone does not allow a firm
conclusion about the existence of the low-mass peak at $1.25 M_{\odot}$, since
they focus just on the mean values and dispersions. In all cases the
relatively high values of the latter dispersions seem to be a consequence
of this methodology.

\vspace{0.4cm}

\noindent{\bf Acknowledgments}

\vspace{0.4cm}

We would like to acknowledge the financial support of the FAPESP
(S\~ao Paulo) Agency and CNPq (Brazil). Drs. M.G. Pereira and F.
D'Amico are gratefully acknowledged for their scientific advice
during this work. An anonymous referee is acknowledged for useful 
suggestions and criticisms that improved the first version of this work.

\vspace{0.4cm}

\noindent{\bf References}

\vspace{0.4cm}

\bigskip Barziv, O., Karper, L., van Kerkwijk, M.H., Telging, J.H., van Paradijs, J. (b), 2001, A\&A, 377, 925

\bigskip Bassa, C.G., van Kerkwijk, M.H., Koester, D., Verbunt, F. (O), 2006, A\&A 456, 295

\bigskip Bhat, N.D.R., Bailes, M., Verbiest, J.P.W. (j), 2008, Phys. Rev. D 77, 124017

\bigskip Casares, J., Gonz\'alez Hern\'andez, J.I., Israelian, G., Rebolo, R. (d), 2010, MNRAS, 401, 2517

\bigskip Champion, D.J., Lorimer, D.R., McLaughlin, M.A., Xilouris, K.M., Arzoumanian, Z., Freire, P.C.C.,
Lommen, A.N., Cordes, J.M., Camilo, F. (z), 2005, MNRAS, 363, 929

\bigskip Clark, J.S. et al.(a), 2002, A\&A, 392, 909

\bigskip Corongiu, A., Kramer, M., Lyne, A.G., Lohmer, O., D'Amico, N., Possenti, A. (A), 2004, Mem. S. A. It. Suppl.
5, 188

\bigskip Demorest, P., Pennucci, T., Ransom, S., Roberts, M., Hessels, J. 2010, Nature, 467, 1081

\bigskip Ferdman, R.D. (J), 2008, Ph.D. thesis, Univ. of British Columbia

\bigskip Finn, L.S., 1994, Phys. Rev. Lett., 73, 1858

\bigskip Freire, P.C.C.(N), 2009, talk given at the Hirschegg 2009:Nuclear Matter at High Density Workshop,
available at http://crunch.ikp.physik.tu-darmstadt.de/nhc/pages/events/hirschegg/2009/prog-2009.html

\bigskip Freire, P.C.C., Camilo, F., Kramer, M., Lorimer, D.R., Lyne, A.G., Manchester, R.N., D'Amico, N. (w),
2003, MNRAS, 340, 1359

\bigskip Freire, P.C.C., Ransom, S.M., Gupta, Y. (D), 2007,  ApJ 662, 1177

\bigskip Freire, P.C.C., Wolszcan, A., van den Berg, M., Hessels, J.W.T (l), 2008a, ApJ 679, 1433

\bigskip Freire, P.C.C., Ransom, S.M., B\'egin, S., Stairs, I.H, Hessels, J.W.T., Frey, L.H., Camilo, F. (H),
2008b, ApJ, 675, 670

\bigskip Freire, P.C.C., Jacoby, B.A., Bailes, M. (L), 2008, in 40 YEARS OF PULSARS: Millisecond Pulsars, Magnetars
and More, AIP Conference Proceedings 983, 488

\bigskip Gelino, D.M., Tomsick, J.A., Heindl, W.A. (I), 2003, Bull. Am. Astron. Soc., 34, 1199

\bigskip Hotan, A.W., Bailes, M., Ord, S.M. (u), 2006, MNRAS 369, 1502

\bigskip Jacoby, B.A., Cameron, P.B., Jenet, F.A., Anderson, S.B., Murty, R.N., Kulkarni, S.R. (x), 2006, ApJL, 644, L113

\bigskip Janssen, G.H., Stappers, B.W., Kramer, M., Nice, D.J., Jessner, A., Cognard, I., Purver, M.B.A (C), 2008,
A\&A, 490, 753

\bigskip Jonker, P.G., van der Klis, M., Groot, P.J. (n), 2003, MNRAS, 339, 663

\bigskip Kramer, M., Stairs, I.H., Manchester, R.H., McLaughlin, M.A., Lyne, A.G., Ferdman, R.D., Burgay, M.,
Lorimer, D.R., Possenti, A., D'Amico, N., Sarkisian, J.M., Hobbs, G.B., Reynolds, J.E., Freire, P.C.C., Camilo, F.(i),
2006, Science 314, 97

\bigskip Kiziltan, B., Kottas, A., Thorsett, S.E., arXiv:1011.4291v1 (2010)

\bigskip Lange, Ch., Camilo, F., Wex, N., Kramer, M., Backer, D.C., Lyne, A.G., Doroshenko, O. (m), 2001, MNRAS
326, 274

\bigskip Lorimer, D.R., Stairs, I. H.; Freire, P. C.; Cordes, J. M.; Camilo, F.; Faulkner, A. J.; Lyne, A. G.; Nice, D. J.; Ransom, S. M.;
Arzoumanian, Z.; Manchester, R. N.; Champion, D. J.; van Leeuwen, J.; Mclaughlin, M. A.; Ramachandran, R.; Hessels, J. W.; Vlemmings, W.;
Deshpande, A. A.; Bhat, N. D.; Chatterjee, S.; Han, J. L.; Gaensler, B. M.; Kasian, L.; Deneva, J. S.; Reid, B.;
Lazio, T. J.; Kaspi, V. M.; Crawford, F.; Lommen, A. N.; Backer, D. C.; Kramer, M.; Stappers, B. W.; Hobbs, G. B.;
Possenti, A.; D'Amico, N.; Burgay, M. (B), 2006, ApJ, 640, 428

\bigskip Mason, A.B., Norton, A.J., Clark, J.S., Negueruela, I., Roche, P. (f), 2010, A\&A, 509, 79

\bigskip Nice, D.H.(M), 2004, in Proceedings of the IAU Symposium 218: Young Neutron Stars and their Environment,
eds. F.M. Camilo \& B.M. Gaensler, Ast. Soc. Pac., San Francisco

\bigskip Nice, D.H., Splaver, E.M., Stairs, I.H. (h), 2001, ApJ 549, 516

\bigskip Nice, D.J., Splaver, E.M., Stairs, I. H. (g), 2003, in Radio Pulsars,
ed. M. Bailes, D. J. Nice, and S. E. Thorsett (Ast. Soc. Pac. 302,
San Francisco), 75

\bigskip Nice, D.J., Stairs, I.H., Kasian, L.E. (y), 2008, in 40 YEARS OF PULSARS: Millisecond Pulsars, Magnetars
and More, AIP Conference Proceedings 983, 453

\bigskip Nomoto, K., 1987, ApJ, 322, 20

\bigskip Nomoto, K., Iben, I., Jr., 1985, ApJ, 297, 531

\bigskip Nomoto, K., Kondo, Y., 1991, ApJ, 367, L19

\bigskip Podsiadlowski, Ph., Langer, N., Poelarends, A. J. T., Rappaport, S., Heger, A.,
Pfahl, E., 2004, ApJ, 612, 1044

\bigskip Podsiadlowski, Ph., Dewi, J. D. M., Lesaffre, P., Miller, J. C., Newton, W. G.,
Stone, J. R., 2005, MNRAS, 361, 1243

\bigskip Poelarends, A. J. T., Herwig, F., Langer, N., Heger, A., 2008, ApJ, 675, 614

\bigskip Quaintrell, H., Norton, A.J., Ash, T.D.C., Roche, P., Willems, B., Bedding, T.R., Baldry, I.K., Fender, R.P. (c),
2003, A\&A, 401, 303

\bigskip Ransom, S.M., Hessels, J.W.T., Stairs, I.H., Freire, P.C., Camilo, F., Kaspi, V.M., Kaplan, K.L. (t),
2005, Science 307, 892

\bigskip Schwab, J., Podsiadwolski, Ph., Rappaport, S., 2010, ApJ, 719, 722

\bigskip Siess, L., 2007, A\&A, 476, 893

\bigskip Splaver, E.M., Nice, D.J., Stairs, I.H., Andrea, A.N., Backer, D.C. (r), 2005, ApJ 415, 405

\bigskip Stairs, I.H., Thorsett, S. E., Taylor, J.H., Wolszczan, A. (K), 2002, ApJ, 581, 501

\bigskip Steeghs, D., Jonker, P.G. (F), 2007, ApJL, 669, L85

\bigskip Steiner, A.W., Lattimer, J.M., Brown, E.F., 2010, ApJ, 722, 33

\bigskip Thorsett, S. E., Chakrabarty, D. (e), 1999, ApJ, 512, 288

\bigskip Timmes, F. X., Woosley, S. E., Weaver, T.A., 1996, ApJ 457, 834

\bigskip van den Heuvel, E. P. J. 2004, in ESA SP 552, 5th INTEGRAL Workshop
on the INTEGRAL Universe, eds. V. Schoenfelder, G. Lichti, \& C. Winkler
(Noordwijk: ESA), 185

\bigskip van den Heuvel, E. P. J. 2010, in , Astrophysics of Neutron Stars 2010, ed. E. Gogus et al.
(AIP Conference Proceedings), in the press.

\bigskip van der Meer, A., Kaper, L., van Kerkwijk, M.H., Heermskerk, M.H.M., van den Huevel, E.P.J. (E), 2007,
A\&A, 473, 523

\bigskip van Kerkwijk, M.H., van Paradijs, J.,  Zuiderwijk, E.J. (k), 1995, A\&A 303, 497

\bigskip Verbiest, J.P.W., Bailes, M., van Straten, W., Hobbs, G.B., Edwards, R.T., Manchester, R.N.,
Bhat, N.D.R., Sarkissian, J.M., Jacoby, B.A., Kulkarni, S.R. (p), 2008, ApJ 679, 675

\bigskip Weisberg, J.M., Taylor,J.H. (q), 2005, in  ASP Conf. Ser. 328, Binary Radio Pulsars, eds. F. A. Rasio \& I. H.
Stairs, Ast. Soc. Pac., San Francisco, 25

\bigskip Zhang, C.M., Wang, J., Zhao, Y.H. Yin, H.X., Song, L.M., Menezes, D.P., Wickramasinghe, D.T.,
Ferrario, L., Chardonnet, P., 2010, A\&A, in the press.

\end{document}